\documentclass[aps,pra,reprint,superscriptaddress]{revtex4-1}

\usepackage[version=3]{mhchem}
\usepackage{amsmath}
\usepackage{pgfplots}
\usepackage{float}
\usepackage[hidelinks]{hyperref}

\usepackage{enumerate}
\usepackage{footnote}
\usepackage{perpage} 
\MakePerPage{footnote} 
\usepackage{booktabs}
\usepackage{changepage} 
\usepackage{siunitx}
\usepackage{systeme,mathtools}
\usepackage{verbatim}
\usepackage[capitalize]{cleveref}

\begin{document}


\title{Theory of transport and conversion in bioelectrochemical systems}



\author{A.C.L de Lichtervelde}
\affiliation{Department of Environmental Technology, Wageningen University, Bornse Weilanden 9, 6708 WG Wageningen, The Netherlands}

\author{A. Ter Heijne}
\affiliation{Department of Environmental Technology, Wageningen University, Bornse Weilanden 9, 6708 WG Wageningen, The Netherlands}

\author{H.V.M. Hamelers}
\affiliation{Wetsus, European Centre of Excellence for Sustainable Water Technology, Oostergoweg 9, 8911 MA Leeuwarden, The Netherlands}

\author{P.M. Biesheuvel}
\affiliation{Wetsus, European Centre of Excellence for Sustainable Water Technology, Oostergoweg 9, 8911 MA Leeuwarden, The Netherlands}

\author{J.E. Dykstra}
\affiliation{Department of Environmental Technology, Wageningen University, Bornse Weilanden 9, 6708 WG Wageningen, The Netherlands}

\date{\today}

\begin{abstract}
Bioelectrochemical systems are electrochemical cells that rely on conductive biofilms covering an electrode. We consider the example of a microbial fuel cell, and we derive a dynamic model of ion transport, biochemical reactions and electron transport inside such a biofilm. After validating the model against data, we evaluate model output to obtain an understanding of the transport of ions and electrons through a current-producing biofilm. For a system fed with a typical wastewater stream containing organic molecules and producing 5 A m$^{-2}$, our model predicts that transport of the organic molecules is not a limiting factor. However, the pH deep within the biofilm drops significantly, which can inhibit current production of such biofilms. Our results suggest that the electronic conductivity of the biofilm does not limit charge transport significantly, even for a biofilm as thick as 100 $\mu\mathrm{m}$. Our study provides an example of how physics-based modelling helps to understand complex coupled processes in bioelectrochemical systems.
\end{abstract}

\maketitle

\section{Introduction}   
Bioelectrochemical systems (BESs) are electrochemical cells in which bacteria catalyze reactions on an electrode \cite{Hamelers2010}. Examples are the microbial fuel cell (MFC) in which electrical energy is recovered from an aqueous organic stream  \cite{Logan2006,Hamelers2010}, the microbial electrolysis cell (MEC) in which electricity is used for the production of hydrogen~\cite{Logan2008,Jeremiasse2010}, microbial electrosynthesis for the production of long-chain hydrocarbons \citep{Rabaey2010,Molenaar2016}, and microbial corrosion~\cite{Enning2014}. In a BES, bacteria attach themselves directly to the surface of electrodes, forming a porous layer called ``biofilm'', through which ion transport occurs. The biofilms have electron-conductive properties, with bacteria performing long-distance electron transport over tens to hundreds of micrometers, to deliver current to the electrode \cite{malvankar2011tunable,schrott2011electrochemical,Lovley2012,Adhikari2016}. In theory, thicker biofilms can produce more current and chemicals per area of electrode.
\\
To improve our understanding of bioelectrochemical systems, a theoretical model of the processes inside the biofilm is necessary. Over the past decade, several models have been developed to integrate knowledge from experimental research \cite{Marcus2007,Picioreanu2007,Oliveira2013,Korth2015}. Accordingly, models have so far relied upon empirical knowledge to an important extent. An important example is the Monod equation, which can be used to calculate rates of substrate utilization in biological systems. However, the dynamic processes in a BES, including biochemical reactions, ion and charge transport, can be described in a more fundamental way by a dynamic system of PDEs \cite{Chen2010}.
In our paper, we present a physics-based dynamic description of ion transport, bioelectrochemical reactions, and electron transport inside conductive biofilms on electrodes. 
As a specific example of the theory, we focus on a biofilm where dissolved organic matter in the form of acetate is converted into electrical energy, such as in an MFC, and we use our modelling framework to address two important questions in the field of bioelectrochemical systems.  
\\
First, it remains unclear which factors limit the current density in an MFC. For instance, it has been argued that transport of electrons is rate-limiting and therefore is the cause of a maximum biofilm thickness and thus current production \cite{strycharz2011application,Pocaznoi2012,Bond2012}, whereas others argue that electron transport can be achieved over greater distances than the typical biofilm thickness and should not be a significant limitation \cite{malvankar2011tunable,Lovley2012,Yates2015}. Other studies demonstrated that within the biofilm protons accumulate at levels likely to inhibit current production \cite{Franks2009}. Second, a fundamental question relates to anode polarization experiments, in which current density is measured while the anode potential is varied \cite{Logan2006,TerHeijne2008,TerHeijne2015}. In these experiments, it is frequently observed that current and power both peak before decreasing to a steady state value \cite{Watson2011,Win2011,Ieropoulos2010,Pocaznoi2012}. This feature has related to performance  limitations of MFCs and the inaccurate assessment of the maximum power density of an MFC \cite{Watson2011,Win2011}. Interpretations of this overshoot phenomenon vary widely across literature \cite{Pocaznoi2012,Watson2011,Win2011}.
\\
In the present work, we show how a dynamic version of our model captures the phenomenon of current overshoot observed in polarization experiments, which helps us to propose an interpretation. Furthermore, we use the model to identify limiting processes for current production with typical wastewater feed. We show that mass transport of organic molecules (called ``substrate'') through the biofilm is not a limiting factor, but the accumulation of protons deep within a biofilm can be limiting. Finally, based on measured biofilm conductivities \cite{malvankar2011tunable}, our model results indicate that electron transport to the anode is not a rate-limiting factor. 
\\
\section{Theory}\label{sec:theory}
\begin{figure}[H]
\centering
\includegraphics[scale=0.39]{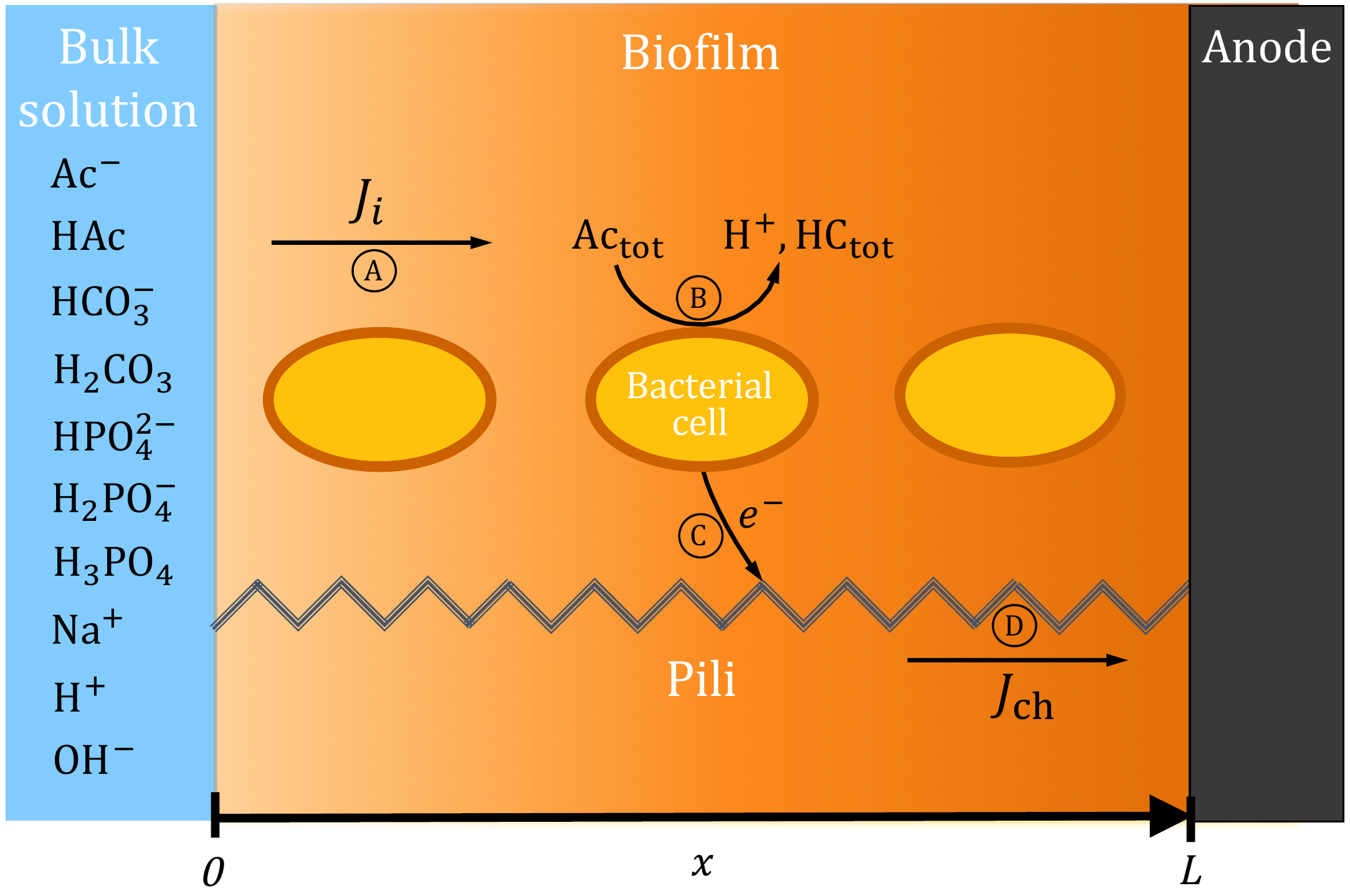}
\caption{Scheme of the coupled biofilm model, which depicts transport and bioelectrochemical reactions in four steps: (A) ion transport, $J_{\mathrm{i}}$, across the biofilm, (B) oxidation of substrate in bacteria, (C) electron transfer to a conductive structure, and (D) charge transport, $J_\mathrm{ch}$, to the anode.}\label{fig:bioanode_scheme}
\end{figure}

To model an electrochemically active biofilm, we present a mathematical framework that consists of four elements, as shown in Fig. \ref{fig:bioanode_scheme}: (A) transport of substrate and products across the biofilm, (B) oxidation of substrate in the bacteria, (C) electron transfer to a conductive structure, e.g. in the form of pili, and (D) charge transport to the anode. 

\subsection{Mass transport across the biofilm}\label{subsec:TRP}
In this work, we focus on transport of ions and other molecules across a one-dimensional planar biofilm, with simultaneous biochemical reactions.  
Biofilms consist of a dispersed phase (the bacterial cells) and a continuous aqueous phase (the extracellular space, abbreviated ``ES''). We define the porosity of the biofilm, $\epsilon$ (m$^{3}$ ES / m$^{3}$ biofilm), as the ratio of the volume of the extracellular space to the total volume of the biofilm and the tortuosity, $\tau$ (m ES / m biofilm), as the ratio of the average transport distance inside the pores to the geometric displacement. Ion transport across the biofilm is described by the Nernst-Planck equation
\begin{equation}
\label{eq:NP}
	J_i =   \: - D_{i,e} \left( \frac{\partial{c_i}}{\partial{x}} + z_i\: c_i\: \frac{\partial{\phi}}{\partial{x}} \right) 
\end{equation}
where $J_i$ is the molar flux of species $i$ (mol m$^{-2}$ s$^{-1}$) across the biofilm, $D_{i,e}$ the effective diffusivity of $i$ in the biofilm (m$^2$ s$^{-1}$), $c_i$ the concentration of $i$ (mol m$^{-3}$ or mM) in the extracellular space, $z_i$ the valence of $i$ (-), $\phi$ the dimensionless electric potential (-) and $x$ the position in the biofilm (m). Potential $\phi$ can be converted to a dimensional voltage by multiplying with the thermal voltage, $V_\mathrm{T} = \frac{R\:T}{F}$, where $F$ is the Faraday constant (96485 C mol$^{-1}$), $R$ the universal gas constant (8.314 J K$^{-1}$ mol$^{-1}$) and $T$ the temperature (298 K). Transport of components in the biofilm is restricted by the presence of the bacteria and their extracellular substances. Therefore, the effective diffusivity in the biofilm is only a fraction of the value in aqueous solution, and is calculated as $D_{i,e} = D_\mathrm{r}\:D_i$, where $D_\mathrm{r} = \frac{\epsilon}{\tau}$ is the relative diffusivity (m$^2$ ES / m$^2$ biofilm) and $D_{i}$ the diffusivity in free solution (m$^2$ s$^{-1}$). Mass conservation holds everywhere in the extracellular space, and is given by
\begin{equation}
\epsilon \frac{\partial c_i}{\partial t} = - \frac{\partial J_i}{\partial x} + r_i + \gamma_{i}
\label{eq:massbalance}
\end{equation}
where $r_i$ is the formation rate of species $i$ (mM s$^{-1}$) due to biochemical reactions in the biofilm, which are described in more detail in Section \ref{subsec:conversions}, and $\gamma_{i}$ is the formation rate of $i$ due to  acid-base reactions. For inert species that do not undergo any of these reactions, we have $r_i=\gamma_{i}=0$, and for the phosphate system, we have $r_i=0$. (Note that, for simplicity, we only consider transport of chemical species in the extracellular space, which is reflected by the presence of the porosity factor in the \textbf{lhs} of Eq. \eqref{eq:massbalance}).

To describe transport of species that participate in acid-base reactions, we group them in the following way. In our calculation we consider the group 
containing acetate species (\ce{Ac$^-$}) and acetic acid (\ce{HAc}) (together also called ``substrate''); the group 
containing bicarbonate species \mbox{($\ce{HCO3^{-}}$)} and carbonic acid ($\ce{H2CO3}$), which are the products of the biochemical conversions; and the group of phosphate species 
($\ce{HPO4^2-} $, $\ce{H2PO4^-} $ and $\ce{H3PO4}$) which are typically used in MFC experiments to buffer the pH in the biofilm \cite{Borsje2016,Hamelers2011,Strycharz-glaven2014}. Finally, the solution contains protons (\ce{H+}) and hydroxyl ions ($\ce{OH-}$), and additional unreactive cations, which we jointly describe as \ce{Na+}-ions. We neglect other anions. 

For each group of species, we set up a mass balance equation by adding up all balances of all species in a group, and as the summation of the $\gamma$-terms within a group is zero, these $\gamma$-terms cancel each other out, resulting in
\begin{align}
    \epsilon \frac{\partial}{\partial t}[\ce{Ac}_\mathrm{tot}] &= -\frac{\partial}{\partial x} \left(J_{\mathrm{\ce{Ac-}}}+J_{\mathrm{\ce{HAc}}}\right) - r_{\mathrm{a}} \label{eq:resmassbal1} \\
    \epsilon \frac{\partial}{\partial t}[\ce{HC}_\mathrm{tot}] &= -\frac{\partial}{\partial x} \left(J_{\mathrm{\ce{HCO3-}}}+J_{\mathrm{\ce{H2CO3}}}\right) + 2 \: r_{\mathrm{cat}} \label{eq:resmassbal2} \\
    \epsilon \frac{\partial}{\partial t}[\ce{H2P}_\mathrm{tot}] &= -\frac{\partial}{\partial x} \left(J_{\mathrm{\ce{H3PO4}}}+J_{\mathrm{\ce{H2PO4-}}}+J_{\mathrm{\ce{HPO4^2-}}}\right) \label{eq:resmassbal3}
\end{align}
where we use the notation [$i$] for concentration (equivalent to the symbol $c_i$), again with dimension mM, and where $\ce{Ac}_{\mathrm{tot}}$ is the total concentration of acetate species, $\ce{HC}_{\mathrm{tot}}$ of bicarbonate species, and $\ce{H2P}_{\mathrm{tot}}$ of phosphate species. These concentrations are given by
\begin{align}
[\ce{Ac}_{\mathrm{tot}}]&=[\mathrm{HAc}]+[\mathrm{\ce{Ac-}}] \\
[\ce{HC}_{\mathrm{tot}}]&=[\mathrm{\ce{H2CO3}}]+[\mathrm{\ce{HCO3-}}] \\
[\ce{H2P}_{\mathrm{tot}}]&=[\mathrm{\ce{H3PO4}}]+[\mathrm{\ce{H2PO4^-}}]+[\mathrm{\ce{HPO4^2-}}].
\end{align}
The rate of formation of acetate is given by $-r_\mathrm{a}$ (Eq. \eqref{eq:ra}) and of bicarbonate by $+2\:r_\mathrm{cat}$ (Eq. \eqref{eq:rcat}), as discussed in detail in Section \ref{subsec:conversions}.
After setting up these balances, we substitute the acid-base equilibria, as listed in \cref{tab:parameters}, into the mass balances, \cref{eq:resmassbal1,eq:resmassbal2,eq:resmassbal3}, after which only one ``master species'' per group remains to be considered in the numerical code, as described in Refs. \cite{Dykstra2014,Paz-Garcia2015,Dykstra2017a,Oren2018}. In this way, the numerical code is much simplified as kinetic expressions and constants for these acid-base reactions are not considered. Furthermore, as acid-base reactions are fast compared to diffusion, considering these reactions would result in a ``stiff'' set of equations, which is numerically more difficult to solve. Lastly, by grouping ionic species, we do not have to make any assumption when it comes to which ionic species within a group exactly participates in a certain reaction (as also discussed in \citet{Dykstra2014}). E.g., we do not have to make any assumption whether the bacteria consume \ce{HAc} or \ce{Ac-}, and whether they produce \ce{H2CO3} or \ce{HCO3-}. 

Besides the mass balances for each group of species, we consider the charge balance in solution, which describes that, at each position in the biofilm, the divergence of the ionic current is equal to the rate of charge transfer from the solution to the pili, $r_{\mathrm{ch}}$ (mM s$^{-1}$)
\begin{equation}
    \sum_i \left( z_{i} \: \frac{\partial J_{i}}{\partial x} \right) = - r_{\mathrm{ch}}
    \label{eq:divioniccurrent}
\end{equation}
where $i$ runs over all ionic species, including protons and hydroxyl ions. 

The elegance of the use of \cref{eq:divioniccurrent} is that in our code no balance needs to be set up related to the ``alkalinity'', or to protons/hydroxyl ions. This makes the model very transparent to set up. Note that even without explicitly setting up an ``alkalinity'' balance, local production of protons (or hydroxyl ions) is calculated correctly by the model (in our MFC problem, typically protons are produced), see \citet{Dykstra2014,Paz-Garcia2015,Dykstra2017a,Oren2018}. 

The electroneutrality condition holds everywhere in the biofilm ($x\:\epsilon\:[0,L]$, where $L$ is the thickness of the biofilm)
\begin{equation}
\label{eq:EN}
 \sum_i{z_i \: c_i} = 0
\end{equation} 
where $i$ again runs over all ions and where we neglect a possible charge of the bacteria or of the extracellular substances. 
Next, we define boundary conditions. First, concentrations at the bulk-biofilm boundary ($B\!/\!B\!F$) are equal to the fixed bulk concentrations
\begin{equation}
\label{eq:BC0}
	c_{i}\rvert_{B\!/\!B\!F} = c_{\mathrm{B},i}	
\end{equation}
where $c_{\mathrm{B},i}$ is the bulk concentration of species $i$. We neglect any bacterial charge in the electroneutrality balance, and thus a Donnan potential drop does not have to be considered across the $B/BF$ interface. This is different from models for ion-exchange membranes or other membranes with charged nanopores. 
Secondly, at the biofilm-electrode boundary ($B\!F\!/\!E$), for each group of ionic species, the sum of the mass fluxes of all species within the group must be zero
\begin{equation}
    \sum_{j} J_{j}\rvert_{B\!F\!/\!E}=0
      \label{eq:BCn}
\end{equation}
where $j$ runs over all ionic species in a group. Eq. \eqref{eq:BCn} also holds for the unreactive cations. Furthermore, at the $B\!F\!/\!E$-boundary the ionic current must be zero
\begin{equation}
    \sum_{i} z_{i} J_{i}\rvert_{B\!F\!/\!E}=0
      \label{eq:BCn2}
\end{equation}
where $i$ runs over all ionic species evaluated in the model. 

\subsection{Oxidation of substrate in the bacteria}\label{subsec:conversions} 
\begin{figure}[H]
\centering
\includegraphics[scale=0.54]{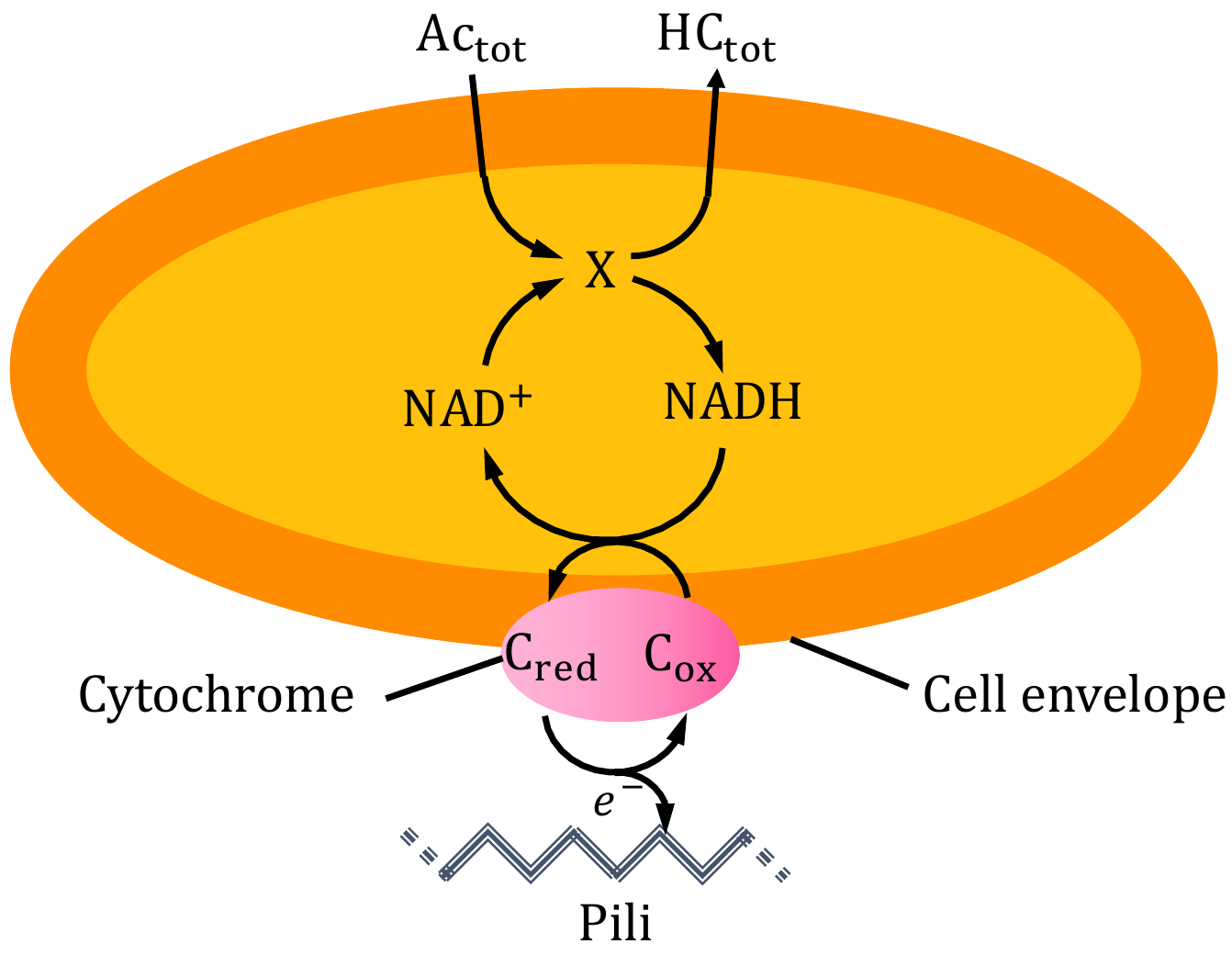}
\caption{Biochemical conversion of substrate and subsequent electron transfer to the matrix of conductive pili. Acetate (Ac$^-$ or HAc) forms an enzymatic complex, X, with NAD$^+$, which is reduced to NADH, while bicarbonate species (\ce{HCO3^-} or \ce{H2CO3}) are produced. NADH reduces a cytochrome on the outside of the cell, which thereafter transfers its electrons to the conductive pili.}
\label{fig:intracellular_reactions}
\end{figure}

We present here a simplified description of the biochemical reactions associated with oxidation of organic matter (acetate) by bacteria under anaerobic conditions (i.e., in the absence of oxygen), see Fig. \ref{fig:intracellular_reactions}.
First, acetate enters the bacterium and forms an enzymatic complex with NAD$^+$, a redox component found in all living cells \cite{nicholls2002bioenergetics,gottschalk2012}, that we use as model electron carrier. Inside the enzyme-substrate complex, that we call ``X'', acetate donates its electrons to NAD$^+$, in a redox reaction yielding NADH (the reduced form of NAD$^+$), which remains inside the cell, and bicarbonate ions and protons, which all leave the cell. 
Next, NADH oxidizes back to NAD$^+$, and transfers its electrons to the cytochromes (proteins that can accept electrons) located on the outer membrane of the cell. Finally, the cytochromes transfer the electrons to conductive pili, which conduct the electrons to the anode. In the following sections, we describe in detail how we model these different steps.

In a bioelectrochemical system, using acetate as a model substrate, the following stoichiometry of conversion is often assumed: \ce{Ac- +4H2O -> 2 HCO3- + 9H+ + 8e-}\cite{Hamelers2010}. However, this stoichiometry is only true in a limited pH range. For this reason, we prefer the numerical approach explained in \cref{subsec:TRP} that eliminates the need to choose for a particular stoichiometry \citep{Dykstra2014}. In our approach, we only have to implement the chemical information that when one acetate species (\ce{Ac-} or \ce{HAc}) is consumed, two bicarbonate species (\ce{HCO3-} or \ce{H2CO3}) form, together with 8 electrons. 

To describe the oxidation of acetate inside the bacteria, we adapt the Butler-Volmer-Monod (BVM) model developed by \citet{Hamelers2011}, which offers a simplified description of the underlying biochemical reactions. We modify the BVM model by 1) assuming that the dissociation of the enzyme-substrate complex is irreversible (we remove the $k_4$ parameter in their model), and 2) adapting it to describe the conversion of acetate to bicarbonate, protons and electrons. This latter modification makes it possible to couple biochemical reactions to the transport of ions.  
With this approach, the oxidation of one acetate species (the Ac$^{-}$ ion or the HAc neutral species) is coupled to the reduction of 4 NAD electron carriers by the formation of an enzyme-substrate complex X, according to 
\begin{equation}
\ce{Ac_{\mathrm{tot}} + 4NAD+ <=>[r_{\mathrm{a}}] 4 X}
\label{ce:bioox}
\end{equation}
while the dissociation of \ce{X} is described by
\begin{equation}
\ce{4X ->[r_{\mathrm{cat}}] 2HC_{\mathrm{tot}} + 4NADH}.\label{ce:bioox_cat}
\end{equation}
In \cref{ce:bioox_cat} the rate of association of $\ce{Ac}_{\mathrm{tot}}$ and \ce{NAD+} into X is given by $r_\mathrm{a}$, and the rate of dissociation of X by $r_\mathrm{cat}$ (both in mM s$^{-1}$). Note that \cref{ce:bioox,ce:bioox_cat} do not take into account charge neutrality, nor atom balance. 

Whereas other BES models rely on empirical expressions such as the double-Monod equation to calculate reaction rates, in our approach, rates of reactions \eqref{ce:bioox} and \eqref{ce:bioox_cat} are proportional to the product of the concentrations of the reactants, following first-order kinetics
\begin{align}
r_{\mathrm{a}} &=k_{\mathrm{a}}\: \ce{[Ac]_{\mathrm{tot}}}\:\ce{[NAD^+]} - k_{\mathrm{d}}\:\ce{[X]} \label{eq:ra} \\
r_{\mathrm{cat}}&=k_{\mathrm{cat}}\:\ce{[X]}\label{eq:rcat}.
\end{align}
\begin{align}
r_{{i}} &=k_{\mathrm{1}}\: \prod_{i=1}^{m}\:c_i\: - k_{\mathrm{2}}\:\prod_{j=1}^{n}\:c_j \\
\end{align}
The rate constant for association is denoted by $k_{\mathrm{a}}$ (mM$^{-1}$ s$^{-1}$), for dissociation by $k_{\mathrm{d}}$ (s$^{-1}$), and for ``catalysis'' by $k_{\mathrm{cat}}$ (s$^{-1}$). 
Note that the theory can also be used to model a biofilm on a cathode by reversing the reaction scheme (the \textbf{lhs} of Eq. \eqref{ce:bioox} becomes the \textbf{rhs} of Eq. \eqref{ce:bioox_cat} and vice-versa). 
Consequently, Eq. \eqref{eq:ra} is replaced by $r_{\mathrm{a}} \:=\: k_{\mathrm{a}}\: $[HC]$_{\mathrm{tot}}\:$[NADH]$\:-\:k_{\mathrm{d}}\:$[X]. No other modification to the theory is required. 
The three kinetic constants of reactions \eqref{ce:bioox} and \eqref{ce:bioox_cat}, $k_\mathrm{cat}$, $k_\mathrm{a}$ and $k_\mathrm{d}$, are related to the classical substrate affinity constant, $K_{\mathrm{s}}$, by
\begin{equation}
K_\mathrm{S} = \frac{k_\mathrm{d}+k_\mathrm{cat}}{k_{\mathrm{a}}}, 
\end{equation}\label{eq:K_S}
which is also called the Michaelis-Menten constant. 
Next, we express the change in concentration of redox complex X as the difference between the  formation rate, $r_{\mathrm{a}}$, and the conversion rate, $r_{\mathrm{cat}}$,
\begin{equation}
\frac{1}{4}\frac{\partial{[\mathrm{X}]}}{\partial{t}} = r_{\mathrm{a}} - r_{\mathrm{cat}}.\label{eq:ODE_Ac-NAD}
\end{equation}

Finally, because the NAD electron carriers do not leave the bacteria, the total concentration, $\ce{[NAD]_{\mathrm{tot}}}$, is position-invariant and constant over time, and is equal to
\begin{equation}
\ce{[NAD]_{\mathrm{tot}}} =  \ce{[NAD+]}+\ce{[NADH]}+\ce{[X]}.\label{NT}
\end{equation}

\subsection{Electron transfer to the matrix of conductive pili}

\subsubsection{Intracellular electron transfer to outer-membrane cytochromes}\label{int_ET}
The mechanism by which electrons are exchanged between the interior of the cell and the extracellular space involves a cascade of redox proteins, such as cytochromes \cite{Yang2010, Bonanni2012, Santos2015}. 
For simplicity, we model the electron transfer between NADH and the outer-membrane cytochromes as a single step, see \cref{fig:bioanode_scheme,fig:intracellular_reactions} \cite{Bonanni2012,Inoue2010}.
Cytochromes can be in the reduced state, denoted by $\ce{C_{\mathrm{red}}}$, or in the oxidized state, $\mathrm{C}_{\mathrm{ox}}$.
As NADH carries two electrons, the redox reaction between NADH and outer-membrane cytochromes is given by
\begin{equation}\label{ce:NAD}
\ce{NADH + 2C_{\mathrm{ox}} + H+ <=>[K_\mathrm{NAD}] NAD+ + 2C_{\mathrm{red}} }.
\end{equation}
We hypothesize that reaction \eqref{ce:NAD} occurs at a much faster rate than the reaction that produces NADH (reaction \eqref{ce:bioox_cat}) and the one that produces $\ce{C_{\mathrm{ox}}}$ (reaction \eqref{ce:ET to pili}). Reaction \eqref{ce:NAD} is thus not considered as a limiting step and we assume equilibrium, as given by
\begin{equation}\label{eq:Km}
K_{\mathrm{NAD}} = \frac{\ce{[NAD+]\:[C_{\mathrm{red}}]^2}}{\ce{[NADH]\:[C_{\mathrm{ox}}]^{2}}\:[\ce{H^+_{\mathrm{i}}}]}
\end{equation}
where [H$^+_{\mathrm{i}}$] is the intracellular proton concentration, which we assume to remain constant at a value of 10$^{-4}$ mM (\textit{i.e.} pH 7). A balance in NAD$^+$ is given by 
\begin{equation}
  2\:\frac{\partial \ce{[NAD+]}}{\partial t} =  - 8\: r_{\mathrm{a}} + r_{\mathrm{cyt}}.
    \label{eq:dNADdt}
\end{equation}
where $r_{\mathrm{cyt}}$ is the rate of oxidation of cytochromes, to be discussed in the next section.
\subsubsection{Extracellular charge transfer from cytochromes to the matrix of conductive pili}
We now describe the charge transfer from bacteria to pili. Cytochromes located on the outside of bacteria transfer their electrons directly to the pili at all positions in the biofilm, as represented in \cref{fig:bioanode_scheme,fig:intracellular_reactions}. Like for $\ce{NAD}_{\mathrm{tot}}$, the total concentration of cytochromes, $\ce{[C]_{\mathrm{tot}}}$, is invariant with time and position 
\begin{equation}\label{eq:cytochrome balance}
\ce{[C]_{\mathrm{tot}} = [C]_{\mathrm{red}} + [C]_{\mathrm{ox}}},
\end{equation}
with a balance over $\ce{[C]_{\mathrm{ox}}}$ given by
\begin{equation}
    \frac{\partial \ce{[C]_{\mathrm{ox}}}}{\partial t} =  - r_{\mathrm{cyt}} - r_{\mathrm{ch}}.
    \label{eq:dCoxdt}
\end{equation}
Note that in our model $r_{\mathrm{cyt}}$ is a dummy parameter, which
cancels out after summing up Eqs. \eqref{eq:dNADdt} and \eqref{eq:dCoxdt}.
The charge transfer from cytochromes to pili is described as a single-electron Faradaic reaction 
\begin{equation}\label{ce:ET to pili}
\ce{C_{\mathrm{ox}} + e- + H+ <=>[r_\mathrm{ch}] C_{\mathrm{red}}} 
\end{equation}
where the charge transfer rate, $r_\mathrm{ch}$, relates to the redox reaction of cytochromes according to the Butler-Volmer equation, a standard model for electrochemical kinetics that has been shown to be applicable to redox proteins as well \cite{Armstrong1993},
\begin{equation}
\label{eq:re_kin}
r_{\mathrm{ch}} =  k_{\mathrm{red}}\: \ce{[C]_{\mathrm{ox}}\:[H+]} \: e^{-\alpha\Delta \phi}  - k_{\mathrm{ox}}\:{\ce{[C]_{\mathrm{red}}}}\:e^{(1-\alpha)\Delta \phi}
\end{equation}
where $k_{\mathrm{red}}$ (mM$^{-1}$ s$^{-1}$) is the rate constant of the reduction reaction, $k_{\mathrm{ox}}$ (s$^{-1}$) of the oxidation reaction, where $\alpha$ [-] is the transfer coefficient, and where $\Delta \phi=\phi_{\mathrm{bf}}(x)-\phi(x)$, with    $\phi_{\mathrm{pili}}(x)$ the dimensionless electric potential in the pili, and $\phi(x)$ the potential of the solution.
\subsection{Charge transport to the anode}
Finally, we must describe charge transport in the matrix of conductive pili. The current density, $J_\mathrm{ch}$ (A m$^{-2}$), is proportional to the biofilm's electronic conductivity $\sigma_\mathrm{bf}$ (S cm$^{-1}$) and the gradient of electric potential across the biofilm, as described by Ohm's law
\begin{equation}\label{eq:Ohm}
    J_{\mathrm{ch}}  = - \sigma_{\mathrm{bf}}\:V_T\:\frac{\partial{\phi_{\mathrm{pili}}}}{\partial{x}}.
\end{equation}
Assuming that pili are not capacitive, no charge accumulates inside. Charge conservation in pili thus implies
\begin{equation}\label{eq:IB_x'}
	 \frac{\partial{J_{\mathrm{ch}}}}{\partial{x}} = r_{\mathrm{ch}}\:F.
\end{equation}
By substituting Eq. (\ref{eq:Ohm}) into Eq. (\ref{eq:IB_x'}) we obtain a final equation for current transfer across the biofilm
\begin{equation}\label{eq:conduction}
 \sigma_{\mathrm{bf}} \: V_T\frac{\partial^2{\phi_{\mathrm{pili}}}}{\partial{x^2}}  = - \:r_{\mathrm{ch}}\:F.
\end{equation}
At the bulk-biofilm boundary ($B\!/\!B\!F$), no current can pass ($J_\mathrm{ch} = 0$), and thus
\begin{equation}
 \frac{\partial{\phi_{\mathrm{pili}}}}{\partial{x}}\bigg\rvert_{B\!/\!B\!F}  = 0.
\end{equation}
We define the anode overpotential $\eta$ (V) as the potential in the pili at the biofilm/electrode boundary minus that in the continuous phase at the bulk-biofilm boundary
\begin{equation}
    \eta= V_T\:(\phi_{\mathrm{pili}}\rvert_{B\!F\!/\!E} - \phi\rvert_{B\!/\!B\!F}).
\end{equation}

Electrons transferred from cytochromes to pili eventually leave the biofilm at the anode surface. The current density there is obtained by integrating the charge transfer rate, $r_{\mathrm{ch}}$, over the biofilm thickness
\begin{equation}
I=-F\:\int_0^{L}{r_{\mathrm{ch}}(x)}dx\label{eq:I(re)}
\end{equation}
where $I$ is the current density at the anode (A m$^{-2}$). 

\begin{table*}[!hp]
\caption{Parameters used in the bioanode model.}
\begin{ruledtabular}   
\begin{tabular}{@{}l l l l r @{}} 
\toprule
\multicolumn{5}{c}{Biofilm parameters}\\
\hline
$L$ &Thickness& 50\footnote{Used in Fig. \ref{fig:Pol_fit}A and Fig. \ref{fig:I_OVS}.}$^{,}$\footnote{Used in Fig. \ref{fig:Pol_fit}B.} / 100\footnote{Used in Fig. \ref{fig:Na+Ac_tot+HC_tot+H2P_tot_ss} and Fig. \ref{fig:5}.}$^{,}$\footnote{Used in Fig. \ref{fig:re_norm_rho}.} & $\mu$m & \cite{franks2010} \\
$\epsilon$ &Mean porosity & 0.9 & m$^3$ ES m$^{-3}$ BF & \cite{Zhang1994}\\
$\tau$ &Mean tortuosity & 2.3 & m ES m$^{-1}$ BF & \cite{Renslow2014}\footnote{Calculated as $\tau =\frac{\epsilon}{D_{\mathrm{r}}}$, with $D_r = 0.4$ \cite{Renslow2014}.}\\
$K_\mathrm{S}$          & Substrate affinity constant & 2.2  & mM & \cite{Hamelers2011}  \\
$k_\mathrm{d}$           & Rate constant for dissociation & 0.1 & s$^{-1}$   & non-sensitive\footnote{The value of this parameter has no significant effect on results, as long as it is positive.}\\
$I_{\mathrm{max}}$     &Maximal current density & 3.1$^\mathrm{a}$ / 2.6$^\mathrm{b}$ / 7.2$^\mathrm{c,d}$& A m$^{-2}$ &fit / fit / - \\
$K_{\mathrm{NAD}}$          & Equilibrium constant for reaction \eqref{ce:NAD} & 5000$^\mathrm{a,c,d}$/500$^\mathrm{b}$ & mM$^{-1}$ &fit \\
$[\ce{NAD}]_{\mathrm{tot}}$ &Total concentration of NAD/NADH components &1     &  mM & estimated from Ref. \cite{belenky2007nicotinamide} \\
$[\ce{C}]_{\mathrm{tot}}$      &Total concentration of cytochromes & 0.8 &mM& close to Ref. \cite{Picioreanu2007} \\
$k_{\mathrm{red}}$ & Electron transfer constant for reduction & 3.1$^\mathrm{a,c,d}$ / 4.3 $^\mathrm{b}$   & mM$^{-1}$s$^{-1}$ & \cite{Ly2013}, fit \\
$k_{\mathrm{ox}}$& Electron transfer constant for oxidation & 0.080$^\mathrm{a,c,d}$ / 0.59 $^\mathrm{b}$  & s$^{-1}$ & \cite{Ly2013}, fit \\
$\alpha$       & Transfer coefficient & 0.5 & $-$ &\cite{Hamelers2011}\\
$\sigma_{\mathrm{bf}}$ & Biofilm conductivity &5 / variable & mS cm$^{-1}$&\cite{malvankar2011tunable}\\
$\eta$ & Overpotential & variable $^\mathrm{a,b}$ / 0.4$^\mathrm{c}$ / 0.25$^\mathrm{d}$ &V& -\\

\tabularnewline
\multicolumn{5}{l}{with $k_{\mathrm{red}} = k^0_{\mathrm{red}}\: e^{+\alpha \phi_{\mathrm{cyt}}} $, $k_{\mathrm{ox}} =k^0_{\mathrm{ox}} \:e^{-(1-\alpha) \phi_{\mathrm{cyt}}} $, $k^0_{\mathrm{red}}= 0.5$ mM$^{-1}$s$^{-1}$ and $ k^0_{\mathrm{ox}} =0.5$ s$^{-1}$, and where $\phi_{\mathrm{cyt}}$ is the formal}\tabularnewline
\multicolumn{5}{l}{potential of outer-membrane cytochromes\footnote{In a polarization curve such as in Fig. \ref{fig:Pol_fit}, $\phi_{\mathrm{cyt}}$ corresponds to the inflection point of the $I-\eta$ curve \cite{Fricke2008}.}, for which we use $\phi_{\mathrm{cyt}} = $3.66$^\mathrm{a,c,d}$ (-0.208 V - SHE) or  4.28$^\mathrm{b}$(-0.192 V - SHE).} \tabularnewline
\tabularnewline
\hline

\tabularnewline
\hline
\multicolumn{5}{c}{Diffusion coefficients in free solution ({*}10\textsuperscript{-9} m\textsuperscript{2} s$^{-1}$)}\\
\hline 
\midrule
$D_{\ce{Ac-}}$  &Acetate& $1.09$&& \cite{vanysek2000ionic} \\
$D_{\ce{HAc}}$  &Acetic acid & $1.30$&&  \cite{vanysek2000ionic}\\
$D_{\ce{H2CO3}}$  &Carbonic acid& $ 1.30$&& \cite{vanysek2000ionic} \\
$D_{\ce{HCO3-}}$  &Bicarbonate& $ 1.18$&& \cite{PHREEQC} \\
$D_{\ce{H+}}$   &Protons& $ 9.31$&& \cite{PHREEQC} \\
$D_{\ce{OH-}}$  &Hydroxide&$ 5.27$&& \cite{PHREEQC} \\
$D_{\ce{Na+}}$  &Sodium& $ 1.33$ && \cite{PHREEQC} \\
$D_{\ce{HPO4^2-}}$  &Hydrogen phosphate&$ 0.69$&& \cite{PHREEQC} \\
$D_{\ce{H2PO4-}}$   &Dihydrogen phosphate&$ 0.85$&& \cite{PHREEQC} \\
$D_{\ce{H3PO4}}$ &Phosphoric acid& $ 1.10 $ && \cite{PHREEQC} \\
\tabularnewline
\tabularnewline
\hline
\multicolumn{5}{c}{Chemical equilibrium constants}\\
\hline
$pK_{\ce{HAc}}$&\ce{CH3COOH  <=> CH3COO- + H+}& $K_\mathrm{HAc} = \frac{[\ce{CH3COO-}][\ce{H+}]}{[\ce{CH3COOH}]}$  &$4.75$ &\cite{stumm2012aquatic}\\
$pK_{\ce{H2CO3}}$&\ce{H2CO3  <=> HCO3- + H+ } & $K_{\ce{H2CO3}} =\frac{[\ce{HCO3-}][\ce{H+}]}{[\ce{H2CO3}]}$&$6.35$ &\cite{stumm2012aquatic}\\
$pK_{\ce{H3PO4}}$&\ce{H3PO4  <=>  H2PO4- + H+}& $K_{\ce{H3PO4}}=  \frac{[\ce{H2PO4-}][\ce{H+}]}{[\ce{H3PO4}]} $& $2.15$ &\cite{stumm2012aquatic}\\
$pK_{\ce{H2PO4-}}$&\ce{H2PO4- <=> HPO4^{2-} + H+  }&$K_{\ce{H2PO4-}} =  \frac{[\ce{HPO4^{2-}}][\ce{H+}]}{[\ce{H2PO4-}]}$&$7.2$ &\cite{stumm2012aquatic}\\
$pK_{\mathrm{W}}$&\ce{H2O  <=> H+ + OH-}  &$K_\mathrm{W}= [\ce{OH-}][\ce{H+}]$&$14$ &\cite{stumm2012aquatic}\\
\tabularnewline
\multicolumn{5}{l}{with $^{10}\mathrm{log}(K_{\mathrm{i}})=3-pK_{\mathrm{i}}$ and $^{10}\mathrm{log}(K_{\mathrm{W}})=6-pK_{\mathrm{W}}$ because in our model concentrations are expressed in mol m\textsuperscript{-3} (mM).}\tabularnewline
\multicolumn{5}{l}{Note that for each species $i$, $K_{i}$ is expressed in mM whereas $K_{\mathrm{W}}$ is in mM\textsuperscript{2}.}\tabularnewline
\tabularnewline
\hline
\multicolumn{5}{c}{Bulk concentrations (mM)}\\
\hline
\ce{[Ac]_{tot}} & Total acetate & 20$^\mathrm{a,b,d}$ / 5$^\mathrm{c}$ &  &\cite{Hamelers2011} / \cite{FAO1992}\\
\ce{[HC]_{tot}} & Total carbonic acid &5$^\mathrm{a,b,d}$ / 2$^\mathrm{c}$  &&\cite{Hamelers2011} / \cite{FAO1992}\\
\ce{[H2P]_{tot}} &Total phosphate &20$^\mathrm{a,b,d}$ / 5$^\mathrm{c}$&&\cite{Hamelers2011} / \cite{FAO1992}\\
pH&&7&&-\\
\tabularnewline
\end{tabular}
\end{ruledtabular}
\label{tab:parameters}
\end{table*}

\section{Results}\label{sec:results}

\subsection{Polarization curves can be fitted with the model}\label{sec:polarization}

\begin{figure}[h]
\includegraphics[scale=0.6]{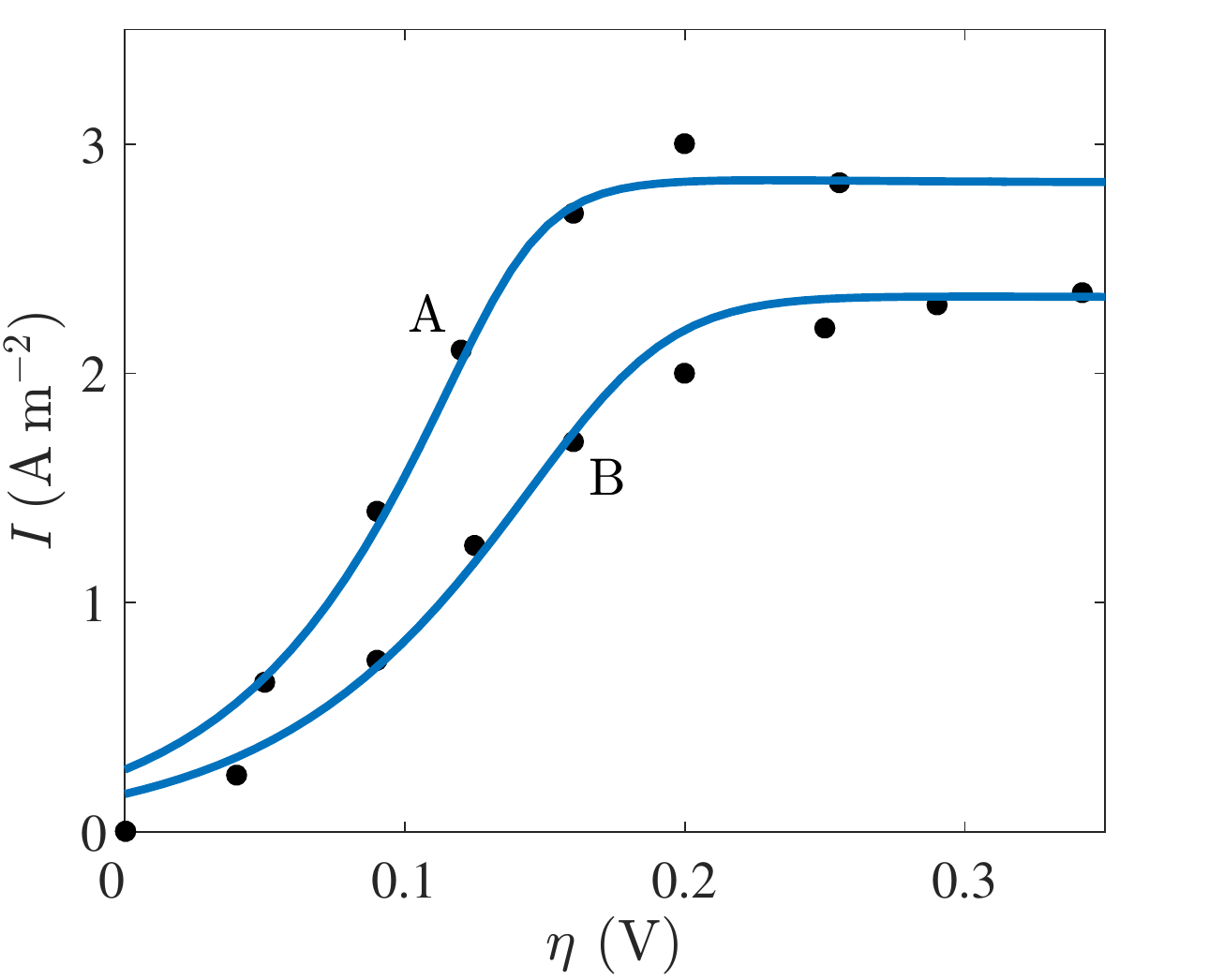}
\caption{Current density, $I$, as a function of anode overpotential, $\eta$. Model calculations (blue curves) fitted to experimental data A and B reported in \citet{Hamelers2011} (black dots).}
\label{fig:Pol_fit}
\end{figure}

Polarization curves show the current density, $I$, as a function of the cell voltage (for a complete electrochemical cell), or for a single electrode, the anode in our case, in function of the electrode (over)potential, $\eta$, see Fig. \ref{fig:Pol_fit}. 

These curves are used to identify conditions for maximum power production, and to calculate energy losses near the electrode interface, see Refs. \citep{TerHeijne2008,Hamelers2011} for a detailed explanation. 

We fit our model to ($I-\eta$)-data reported by \citet{Hamelers2011}, and make an attempt to determine the various unknown parameters of the model: the formal potential of the cytochromes, $\phi_{\mathrm{cyt}}$, the equilibrium constant between NAD and cytochromes, $K_{\mathrm{NAD}}$, the three kinetic constants, $k_{\mathrm{a}}$, $k_{\mathrm{cat}}$ and $k_\mathrm{d}$ (which are related via $K_\mathrm{{S}}$), and in addition we have the total concentration of cytochromes and redox-complexes, $\mathrm{C}_{\mathrm{tot}}$ and $\mathrm{NAD}_{\mathrm{tot}}$. 

Interestingly, we can analyze our model, and find that at steady-state (all time-derivatives are equal to zero) and when [Ac]$_{\mathrm{tot}}$ is relatively high compared to $K_{\mathrm{S}}$, a simple expression holds for the maximum current density, $I_{\mathrm{max}}$, namely \citep{Hamelers2011}
\begin{equation}
I_{\mathrm{max}}= + 8 \: k_{\mathrm{cat}} \ce{[NAD]}_{\mathrm{tot}} \: F \: L \label{eq:Imax}
\end{equation}
and thus we can derive information about [NAD]$_{\mathrm{tot}}$ or $k_{\mathrm{cat}}$ from the measured current density, $I_{\mathrm{max}}$. The validity of this equation is independent of transfer rates of \ce{Ac-} or HAc, or of charge transfer in the pili; it only requires that [Ac]$_{\mathrm{tot}}$ $>>$ $K_S$. 

Based on the limited data available we cannot convincingly derive values for all parameters, so various factors were estimated. The values given to the parameters should be taken as an example, and not as authoritative information. The resulting parameters and their origin are listed in \cref{tab:parameters}. Note that $k_{\mathrm{cat}}$ can be calculated using Eq. \eqref{eq:Imax} from the given value of $I_{\mathrm{max}}$. Similarly, $k_{\mathrm{a}}$ can be derived using Eq. \eqref{eq:K_S} and the given values of $K_{\mathrm{S}}$, $k_{\mathrm{d}}$ and  $k_{\mathrm{cat}}$. 
The model describes the two datasets well, although current density is overestimated at low anode overpotential in both case. 
Note that it is possible that a different parameter set than the one used here leads to better fits of the experimental data.  

At higher $\eta$, the fit becomes more accurate for both datasets, see \cref{fig:Pol_fit}. Interestingly, the values of $\phi_{\mathrm{{cyt}}}$  used in fitting data A and B are close to each other, see \cref{tab:parameters}, suggesting that similar types of cytochromes were active in biofilms A and B. These values are in the range of redox potentials reported in other voltammetry experiments with biofilms on anodes \citep{magnuson2001,Fricke2008}. The value for $I_{\mathrm{max}}$ is higher for polarization curve A (3.1 A m$^{-2}$) than for B (2.6 A m$^{-2}$), indicating a higher metabolic activity or a thicker or denser biofilm.

\subsection{Steady-state concentration profiles across the biofilm} \label{sec:conc_prof_FOP}
\begin{figure}
\centering
\includegraphics[scale=0.6]{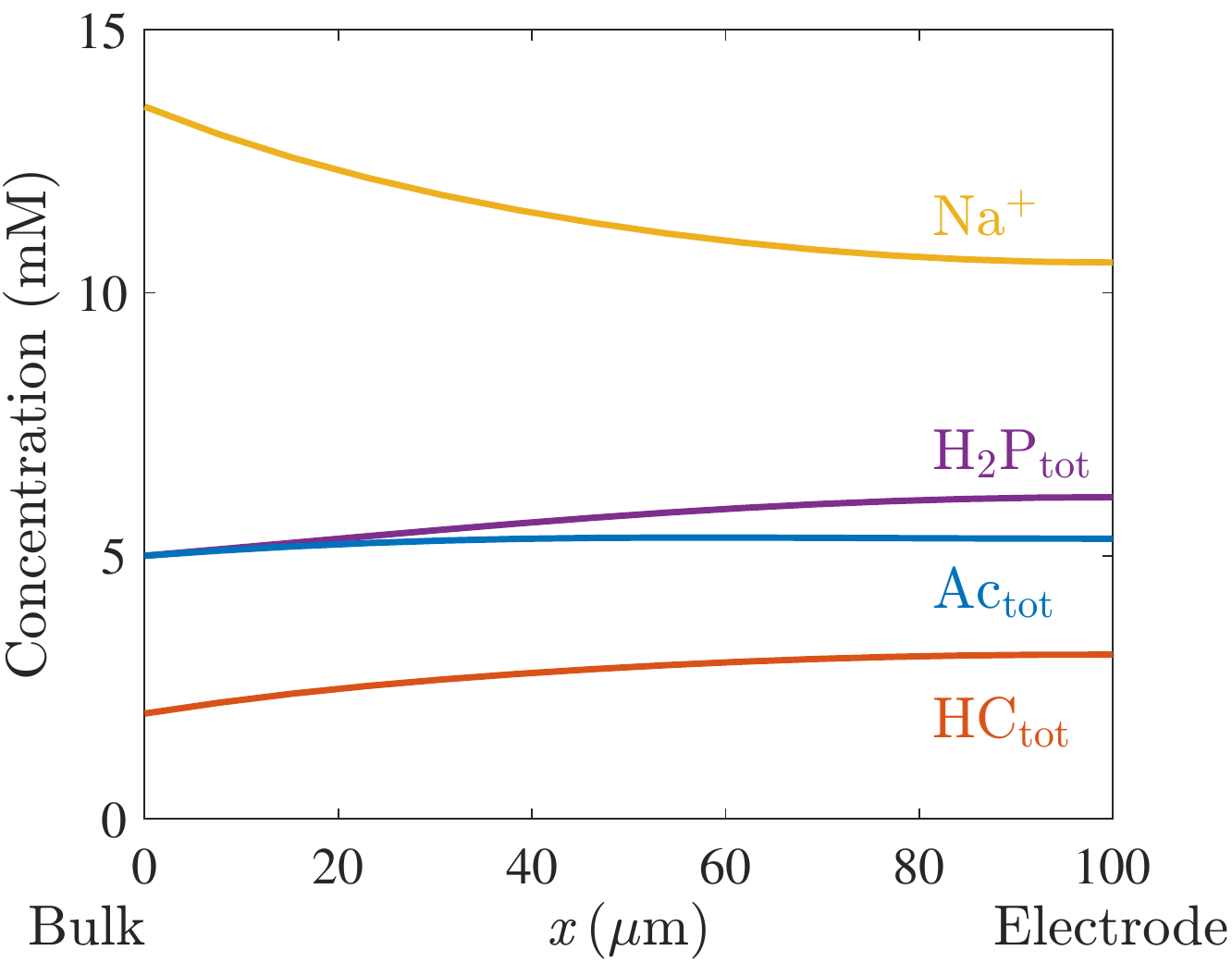}
\caption{Steady-state concentration of sodium, acetate, bicarbonate and phosphate groups in the biofilm.}
\label{fig:Na+Ac_tot+HC_tot+H2P_tot_ss}
\end{figure}

To study mass transfer in the biofilm, steady-state concentration profiles of several chemical species are shown in \cref{fig:Na+Ac_tot+HC_tot+H2P_tot_ss} ($I = \SI{5}{A\:m^{-2}}$).
Whereas previous modelling studies have typically used excess of substrate and buffer, here we evaluate mass transfer in conditions most relevant for MFC operation: the composition of the bulk solution, see \cref{tab:parameters}, is set so to reproduce so-called ``strong domestic wastewater'' with a concentration of Total Dissolved Solids of $\sim$ 1.2 g/L \cite{FAO1992}.

Because of the development of an electric potential gradient across the biofilm,  concentration gradients of ionic species do not necessarily imply there is transport; e.g., in steady-state, there is no transport of \ce{Na+} across the biofilm, but we observe a concentration gradient. The concentration profile of acetate shows accumulation towards the electrode. This is counter-intuitive at first sight, as we would rather expect the concentration of acetate to decrease with biofilm depth due to biochemical conversions \eqref{ce:bioox}. However, as acetate ions experience migrational forces towards the electrode because of the gradient in electrical potential, ions can be transported against their concentration gradient. The almost flat acetate profile suggests that for typical domestic wastewater concentrations \cite{FAO1992}, mass transfer does not restrain the availability of substrate in the biofilm, even in a biofilm of $L=100$ $\mu$m in thickness. 
\begin{figure}[!ht]
\centering\includegraphics[scale=1]{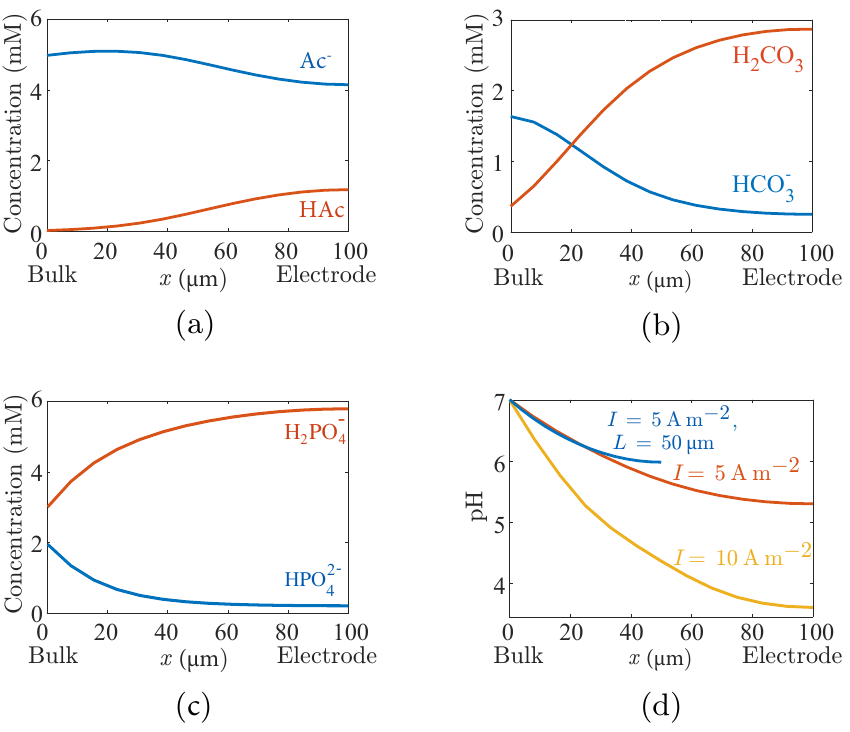}
\caption{Steady-state concentration profiles in the bioanode, for acetate (a), bicarbonate (b) and phosphate species (c), and pH profile for a biofilm with thickness $L=50$ $\mu$m and $I=$ 5 A m$^{-2}$ or 100 $\mu$m and 5 or 10 A m$^{-2}$ (d).}
\label{fig:5}
\end{figure} 

Let us now study concentration profiles of individual ions in the biofilm. Fig \ref{fig:5}a shows that the \ce{Ac-} concentration decreases as function of $x$, whereas \ce{HAc} increases. This is due to the steep pH profile that develops across the biofilm, see Fig. \ref{fig:5}d. Furthermore, Fig. \ref{fig:5}b and c show that the concentrations of \ce{H2CO3} and \ce{H2PO4-} increase with biofilm depth as a result of proton production, whereas the concentrations of \ce{HCO3-} and \ce{HPO4-2} decrease, respectively.

Fig. \ref{fig:5} also shows that the ionic current is mainly carried by the group of ionic species with a pK-value closest to the local pH in the biofilm. Thus, when pH is close to 7.2 ($pK_{\ce{H2PO4-}}$) the current is mainly carried by \ce{HPO4^2-} towards the electrode, and \ce{H2PO4^-} is transported in the reverse direction, resulting in strong concentration gradients of these ionic species as observed in Fig. \ref{fig:5}c; when pH is close to 6.35 ($pK_{\ce{H2CO3}}$) the current is mainly carried by \ce{HCO3-} and \ce{H2CO3}; and when pH is close to 4.75 ($pK_{\ce{HAc}}$) the current is mainly carried by \ce{CH3COO-} and \ce{CH3COOH}.

Fig. \ref{fig:5}d shows the pH profile in the biofilm as function of thickness ($L = 50$ and 100 $\mu$m). We observe that pH drops with increasing depth, most strongly for the thickest biofilm. Besides biofilm thickness, the acidification of the biofilm also depends on current density. 
Indeed, if current density is doubled to $I=10$ A m$^{-2}$, pH at the anode side of the 100 $\mu$m-thick biofilm drops to pH = 3.8. In literature, current production by a biofilm of \textit{Geobacter sulfurreducens} has been reported to drop by \textit{ca}. 50$\%$ when bulk pH dropped from 6.9 to 6.15 (Ref. \cite{Franks2009}, Fig. 7) and to be completely inhibited at pH 5 (Ref. \cite{Patil2011}, Fig. 2b). Thus, our results suggest that the accumulation of protons deep within a biofilm is possibly the most important bottleneck to the increase of current density in bioanodes.

\subsection{Impact of biofilm conductivity on cytochromes and current production}
\begin{figure}
\centering
\includegraphics[width=0.47\textwidth]{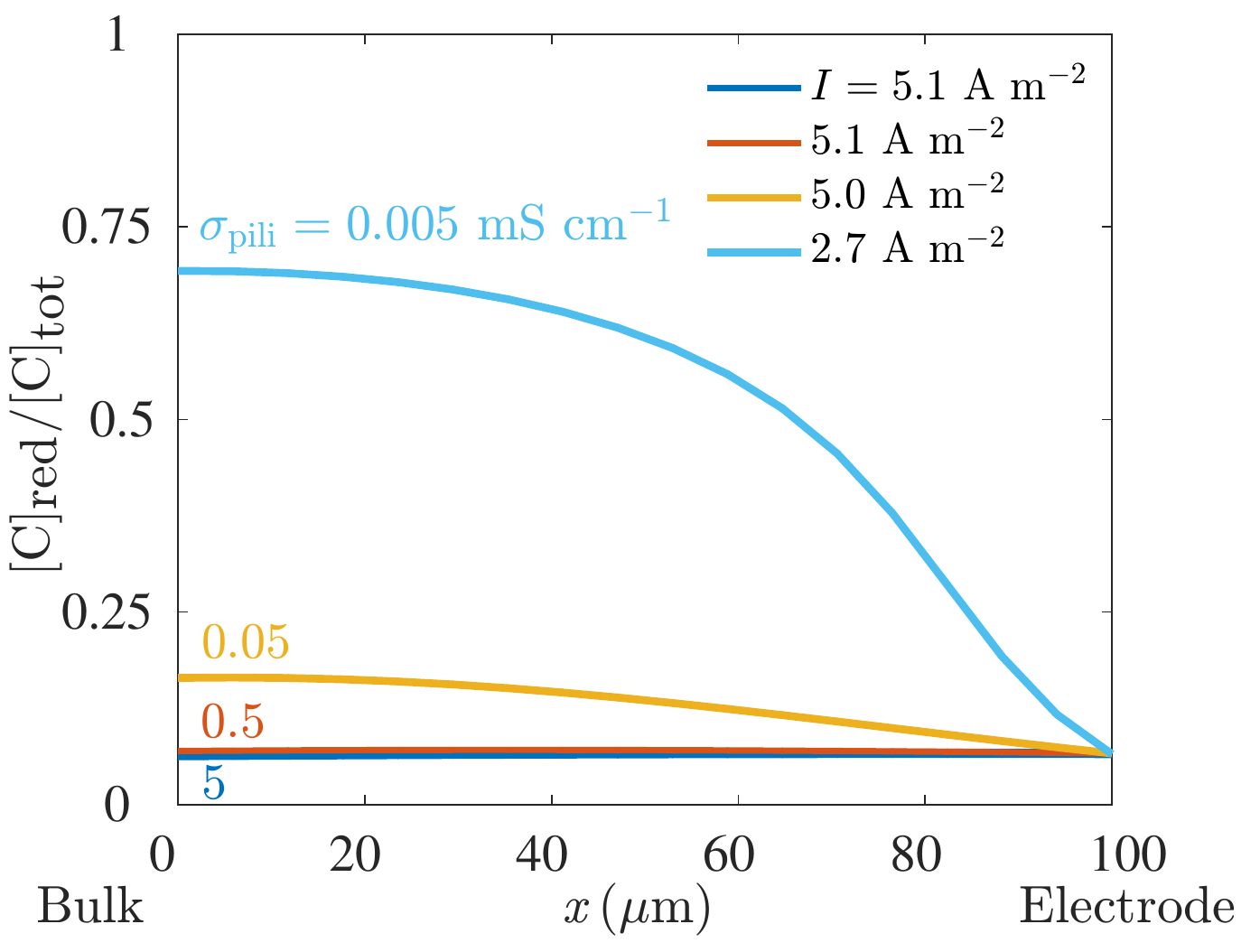}
\caption{Impact of biofilm conductivity, $\sigma_\mathrm{bf}$, on the redox state of cytochromes (expressed as [C]$_\mathrm{red}$/[C]$_\mathrm{tot}$) and current production, $I$ (see legend). As conductivity decreases ($\sigma_\mathrm{bf}$ = 5, 0.5, 0.05 and 0.005 mS cm$^{-1}$), electrons are transported less efficiently towards the anode and accumulate in the form of reduced cytochromes. The development of such a redox gradient across the biofilm decreases current production. However, based on conductivity values reported from experiments (0.5 and 5 mS cm$^{-1}$ \cite{malvankar2011tunable}), no significant redox gradient or decrease in current production is obtained.} 
\label{fig:re_norm_rho}
\end{figure}

The nature of the extracellular electron transport (EET) performed by anode-respiring bacteria such as \textit{Geobacter} and \textit{Shewanella} \textit{sp.} is debated \cite{malvankar2011tunable, strycharz2011application, Snider2012, Bond2012, Yates2015}.\linebreak Most importantly, it remains unclear to what extent EET might be a limiting factor for the development of thick biofilms and for current production. 
In this section we use the model to predict limitations in current production for different values of the electronic conductivity of the biofilm. As a reference, we use conductivity values of $\sigma$ = 0.5 and 5 mS cm$^{-1}$ measured by \citet{malvankar2011tunable} for a 50 $\mu$m-wide gap filled with conductive biofilm of \textit{Geobacter Sulfurreducens} in pure culture. According to Ref. \cite{malvankar2012biofilm}, increasing the width of the gap up to 100 $\mu$m gave similar results. The bioanode is operated at steady state ($I_{\mathrm{max}}$ = 7.2 A m$^{-2}$) and the anode overpotential is set to $\eta=0.25$ V, a value at which maximal current production is reached (as in Fig. \ref{fig:Pol_fit}). Other parameters are listed in \cref{tab:parameters}. 
Fig. \ref{fig:re_norm_rho} shows the concentration of reduced cytochromes, [C]$_\mathrm{red}$ (expressed as a fraction of the total cytochrome concentration, [C]$_\mathrm{tot}$), in function of the position in the biofilm.

In Fig. \ref{fig:re_norm_rho}, we observe that as conductivity is lowered (from 5 to 0.005 mS cm$^{-1}$), electrons are transported less efficiently towards the anode and build up in the form of reduced cytochromes. A gradient of reduced cytochromes develops across the biofilm, and this is associated with a decrease in current production (see legend). 

The bottom blue and red curves corresponds to the two conductivity measured in anode-respiring biofilms \cite{malvankar2011tunable}. For this range of conductivity, [C]$_\mathrm{red}$ remains almost constant across the biofilm layer, meaning that electron transport is very efficient. The current density at the anode is $I =$ 5.1 A m$^{-2}$ in both cases.  
In these two calculations, we have assumed all cells to be well connected to the matrix of conductive pili. Due to the heterogeneous structure of biofilms however, some cells encounter more resistance when transferring electrons to the anode. To account for this eventuality, we consider a significantly lower conductivity of \mbox{$\sigma_{\mathrm{bf}}= $0.05 mS cm$^{-1}$}, \textit{i.e.}, 10 to 100 times lower than the values reported \cite{malvankar2011tunable}. For this value, the concentration of reduced cytochromes starts to increase towards the bulk solution (3$^\mathrm{rd}$, yellow curve), and yet the current density at the anode is still \mbox{5 A m$^{-2}$}. Only when $\sigma_{\mathrm{bf}}$ is further reduced by a factor 10 (to 0.005 mS cm$^{-1}$), an important gradient in [C]$_\mathrm{red}$ appears across the biofilm (top, light-blue curve). Close to the bulk solution, 70$\%$ of cytochromes are in their reduced form, and current production drops to 2.7 A m$^{-2}$. In summary, the fact that lowering experimental values by a factor of 100 to 1000 is needed to obtain significant electron transport limitations suggests that conductivity in the biofilm is not a limiting factor for current production, even for a relatively thick \mbox{($L=$ 100 $\mu$m)} biofilm.

\subsection{Current overshoot in polarization experiments}
\begin{figure}[H]
\centering
\includegraphics[scale=0.6]{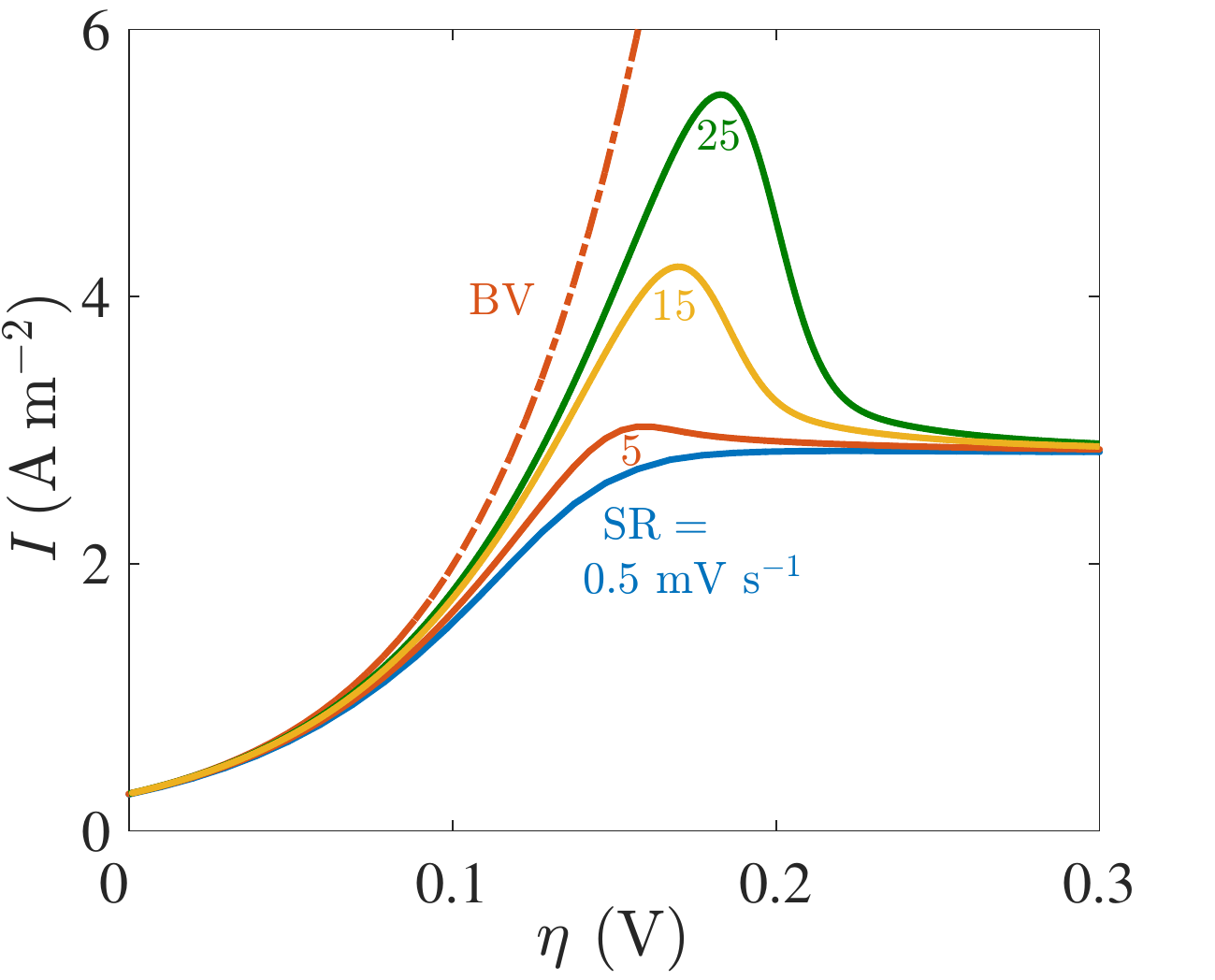}
\caption{Current density, $I$, as a function of anode overpotential, $\eta$, at different scan rates (SR = 0.5, 5, 15 and 25 mV s$^{-1}$). The dashed red curve is the Butler-Volmer (BV) current, obtained from the charge transfer reaction \eqref{ce:ET to pili} with constant cytochrome concentrations (see text). As scan rate is increased, current follows the BV behaviour more closely, resulting in an overshoot that scales with scan rate.}
\label{fig:I_OVS}
\end{figure}

In this final section, we discuss results generated with our model to reproduce the phenomenon of ``current overshoot'' that is frequently observed in polarization experiments, and which is not yet well understood \cite{Watson2011,Ieropoulos2010,Win2011}.
In a polarization experiment the anode potential is gradually changed with a certain scan rate (in mV/s). \mbox{Fig. \ref{fig:I_OVS}} shows calculated polarization curves at different scan rates, and also shows the Butler-Volmer (BV) current, \textit{i.e.} the current predicted by the Butler-Volmer model for an electrochemical reaction with the concentration of reactants kept constant, such that depletion / accumulation of reactants does not affect the predicted current.
In our model we simply calculate the BV current by setting concentrations of cytochromes to their value at zero overpotential ($\mathrm{[C]_{ox}(\eta=0)} \simeq 0 $ mM and $\mathrm{[C]_{red}(\eta=0)} \simeq \mathrm{[C]_{tot}}$). In this case, the BV equation predicts the current to increase exponentially with overpotential, see \cref{fig:I_OVS}. 

In our full model, at a relatively low scan rate of 0.5 mV s$^{-1}$, as used in Fig. \ref{fig:Pol_fit}, current rises smoothly and levels off at a stable value. However, when the scan rate is increased to 5 mV s$^{-1}$, we observe an overshoot in the current. Based on the theory presented here, we interpret this phenomenon as follows: 
in the early part of the experiment ($\eta <$ 0.1 V), the rise of current is dictated by the rate of charge transfer between cytochromes and pili (Eq. \eqref{eq:re_kin}), which increases exponentially with anode potential, following the BV current closely. As current rises further, an increasing number of cytochromes become oxidised, which slows down the rise of current (\ce{[C]_{red}} drops and \ce{[C]_{ox}} increases in Eq. \eqref{eq:re_kin}), and the current no longer follows the BV current. Meanwhile, the oxidation of cytochromes triggers the conversion of substrate to electrons (via the NAD/NADH components, Eq. \eqref{ce:NAD}), reducing \ce{[C]_{ox}} into \ce{[C]_{red}} and allowing current production to be sustained. When finally oxidation and reduction of cytochromes balance each other, the steady-state current is reached. If the scan rate is increased, more cytochromes are oxidized per unit of time, producing a transient current which magnitude increases with scan rate. 
This transient current adds up to the current obtained from substrate conversion and is responsible for the overshoot in current triggered at higher scan rates, as depicted in Fig. \ref{fig:I_OVS}.

\section{Conclusions}
We developed an electrochemical transport model for conductive biofilms on electrodes. 
The model combines mass transport with acid-base equilibria, biochemical reactions and electron transport across the biofilm towards and from the electrode. To describe biochemical reactions inside bacteria, we apply mass action kinetics to an idealized enzymatic reaction, leading to a more general description than the Monod expressions employed in other models. 
As a case study, we derived the model for a microbial fuel cell, however, the theory can be applied to both types of electrodes present in bioelectrochemical systems.
Our calculations show that current production is neither limited by the transport of organic molecules in the biofilm, nor by the biofilm's electronic conductivity. Instead, our calculations predict low pH values deep within the biofilm, which have been associated with inhibition of microbial growth and current production. 

\section*{Acknowledgement}
The research is supported by the Dutch Technology Foundation STW, which is part of the Netherlands Organization for Scientific Research (NWO), and which is partly funded by the Ministry of Economic Affairs (VENI grant no 13631).

%

\end{document}